\documentclass[prb,letterpaper,twocolumn]{revtex4-1}
\usepackage[utf8x]{inputenc}
\usepackage{amsmath}
\usepackage{graphicx}
\usepackage{dcolumn}
\usepackage{bm}
\usepackage{subfigure}
\usepackage{epstopdf}

\begin{document}

\begin{abstract}
BaSn$_2$ has been shown to form as layers of buckled stanene intercalated by barium ions~\cite{Kim_2008}. However, despite an apparently straightforward synthesis and significant interest in stanene as a topological material, BaSn$_2$ has been left largely unexplored, and has only recently been recognized as a potential topological insulator.  Belonging to neither the lead nor bismuth chalcogenide families, it would represent a unique manifestation of the topological insulating phase. Here we present a detailed investigation of BaSn$_2$, using both {\it ab initio} and experimental methods. First-principles calculations demonstrate that this overlooked material is a indeed strong topological insulator with a bulk band gap of 360meV, among the largest observed for topological insulators. We characterize the surface state dependence on termination chemistry, providing guidance for experimental efforts to measure and manipulate its topological properties. Additionally, through {\it ab initio} modeling and synthesis experiments we explore the stability and accessibility of this phase, revealing a complicated phase diagram that indicates a challenging path to obtaining single crystals.
\end{abstract}
\title{\bf BaSn$_2$: A new, wide-gap, strong topological insulator } 
\author{Steve M. Young}
\affiliation{Center for Computational Materials Science, United States Naval Research Laboratory, Washington, D.C. 20375, USA}\author{S. Manni}
\affiliation{Department of Physics and Astronomy, Iowa State University, Ames, IA 50011, USA}
\author{ Junping Shao}
\affiliation{Department of Physics, Applied Physics and Astronomy, Binghamton University, State University of New York, Binghamton, New York 13902-6000, USA}
\author{Paul C. Canfield}
\affiliation{Department of Physics and Astronomy, Iowa State University, Ames, IA 50011, USA}
\affiliation{Ames Laboratory, Iowa State University, Ames, IA 50011, USA}
\author{Aleksey N. Kolmogorov}
\affiliation{Department of Physics, Applied Physics and Astronomy, Binghamton University, State University of New York, Binghamton, New York 13902-6000, USA }

\maketitle
\section{Introduction}
The prediction~\cite{Kane05p146802,Kane05p226801,Fu_2007_2} and realization~\cite{Konig_2007,Hsieh_2008,Hsieh_2009} of topological insulators -- materials with an insulating bulk and robust, metallic surface states -- a decade ago set off an explosion of interest in topological phases of matter~\cite{Hasan_2010,Kong2011,Yan_2012,Ando_2013,Yokoyama2014,Hasan15}.  Since then, new insulating, semimetallic, and metallic phases with nontrivial topology have been proposed and investigated~\cite{futci,Burkov_2011,Fu_2011,Tanaka2012,Yang2012,Young_2012,Dzero_2012,Kim_2013,Steinberg_2014,Rasche_2013,Zaheer_2013,Liu_2014_1,Liu_2014_2,Borisenko_2014,Young_2015,Lv_2015,Kim_2015,Soluyanov_2015,Bansil_2016}.  However, the majority of these efforts has been theoretical. In the case of 3D topological insulators, only a few families of compounds -- all of which are lead and/or bismuth chalcogenides -- with substantial band gaps have been synthesized and experimentally verified~\cite{Hsieh_2009,Lin_2010,Kuroda_2010,Kuroda_2012}, and of those only bismuth selenide has been widely studied. Bi$_2$Se$_3$ appears ideal in theory: it is a simple, binary compound that is easily grown and has a band gap of about 300meV.  However, it is plagued by defects that render the bulk conductive, preventing straightforward utilization and investigation of surface states~\cite{Ando_2013,Li_2014,Li2014,Hsieh2011}.  While these can be ameliorated in principle by careful alloying, implementation of this strategy is still ongoing~\cite{Zhang_2011,Ren_2011,Ren_2012}.  Significant progress remains to be made for topological insulators to be technologically viable.  

\begin{figure}
{
\includegraphics[bb= 0 0  900 400, width=8.5cm]{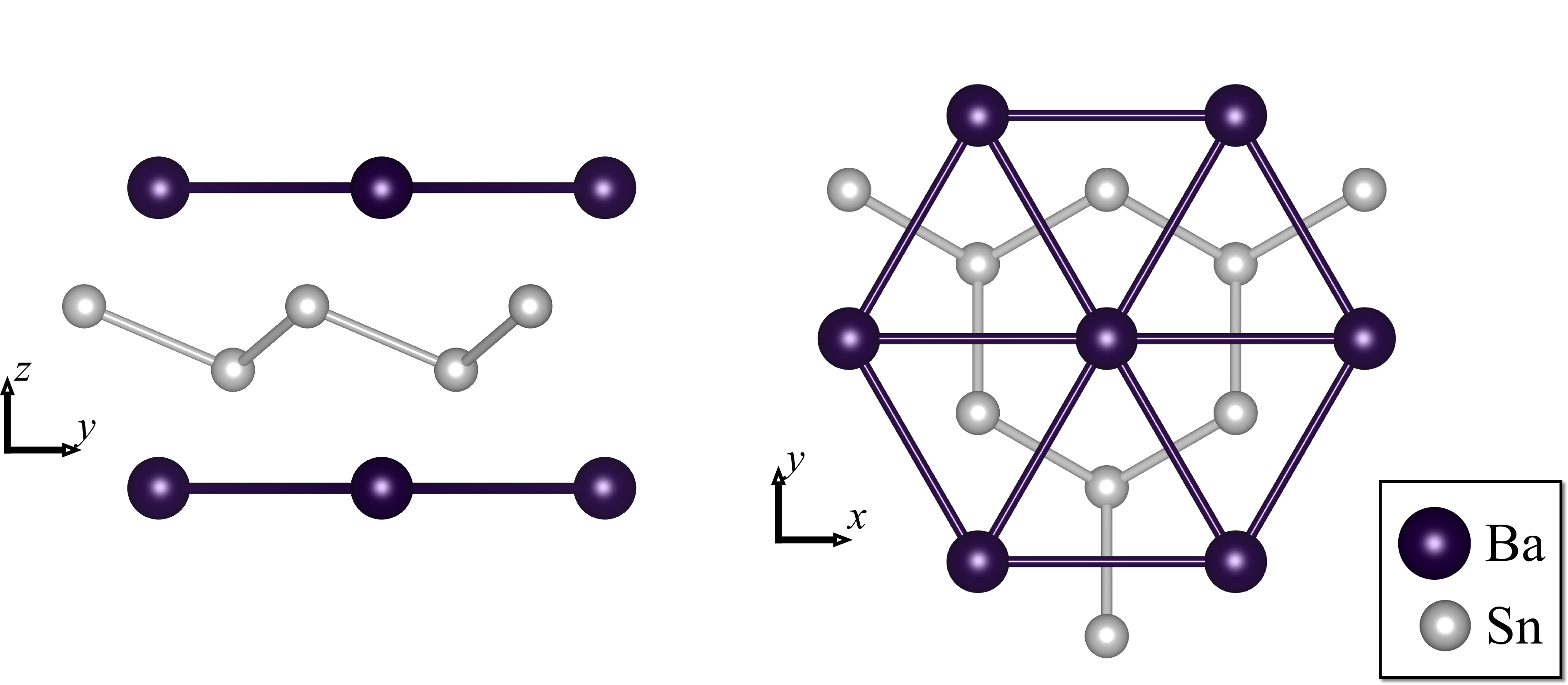}
}

\caption{BaSn$_2$ is composed of alternating layers of tin and barium. The tin layer forms a buckled honeycomb lattice, essentially forming stanene.  Barium forms a flat triangular network, with barium atoms vertically aligned with the hollows of the stanene plaquettes. }\label{fig:struct}
\end{figure}

It is therefore highly advantageous to expand the portfolio of viable compounds.  Here we present a new wide-gap topological insulator, a binary compound that is not a lead or bismuth chalcogenide: BaSn$_2$. Little is known about barium stannide; the published phase diagram for Ba-Sn is incomplete and does not contain information about the stability of binary phases near the BaSn$_2$ composition at high temperatures~\cite{massalski_1990}. The only reported observation of the the compound indicates that it forms from the elements under typical synthesis conditions for Sn-based materials and that it crystallizes in the EuGe$_2$-type hexagonal structure comprising corrugated Sn layers (Fig.~\ref{fig:struct})~\cite{Kim_2008}.   Our recent density functional theory (DFT) investigation of metal distannides showed that a fully connected 3D Sn framework expected to be stable in NaSn$_2$ is effectively broken up by the large Ba atoms into a stacking layers of buckled stanene--itself a two dimensional topological insulator~\cite{Xu_2013,Zhu_2015,Xu15}--in BaSn$_2$~\cite{ak31}. Preliminary calculations of the bulk band structure and $Z_2$ invariants revealed a potential for BaSn$_2$ to display topologically non-trivial behavior\cite{ak31}. Our present DFT calculations with more accurate treatment of the excited states demonstrate that the material is indeed a strong topological insulator and predict a bulk band gap of 360 meV, placing it among the widest known~\cite{Ando_2013,Bansil_2016}. We characterize the electronic structure, including the surface states and their variation with surface chemistry.  Additionally, we explore the surrounding space of the phase diagram experimentally and computationally to reveal synthesis conditions promoting formation of BaSn$_2$ for future investigation of the compound's topological properties. We establish BaSn$_2$ to be a low-temperature phase that does not grow directly from liquid flux.

\section{Methods}
Band structure and surface calculations were performed using ~{\small QUANTUM ESPRESSO}, using PBE~\cite{pbe} and the HSE~\cite{Heyd06p219906} hybrid functional, the latter of which has been demonstrated to offer significantly improved descriptions of semiconductors~\cite{Janesko_2009}.  Bulk calculations were done on a $8\times 8\times 6$ k-point grid; the q-point grid for the exact exchange was $8\times 8\times 6$ as well.  Calculations for slabs of eleven BaSn$_2$ layers were performed at the level of PBE on $8\times8$ k-point grids with at least 10\AA~of vaccuum.  In all cases, the energy cutoff was 50Ry.  HSE band structures were obtained using Wannier interpolation via Wannier90~\cite{WANNIER90}.  
The stability analysis was carried out with {\small VASP}\cite{VASP1,VASP2} at the level of PBE, using the projector augmented wave method~\cite{PAW} and a 500 eV cut-off. Fine meshes were used to ensure numerical convergence of relative energies to within 1-2 meV/atom. Vibrational contributions to free energy at elevated
temperatures were found with the frozen phonon method as implemented
in {\small PHON} \cite{PHON}. As in our previous study of metal
stannides \cite{ak31}, we considered a set of known structures in this
and related $M$-Sn binaries at the ambient pressure and carried out
unconstrained evolutionary searches at select compositions using
{\small MAISE} package \cite{maise}.Structure images were created using {\small VESTA}~\cite{VESTA}.
For the single crystal growth, high purity ($>99.9$\%) Ba and Sn were packed in a three cap Ta-crucible as described elsewhere~\cite{Jesche_2014}, which was then sealed in a silica ampule in partial pressure of Argon and heated up to 950$^\circ$~C followed by slow cooling
down to 680$^\circ$~C, where they were spun in a centrifuge.

\section{Electronic Structure}
\subsection{Bulk}
\begin{figure}
\subfigure[]{\includegraphics[bb= 0 0  576 430,width=8.5cm]{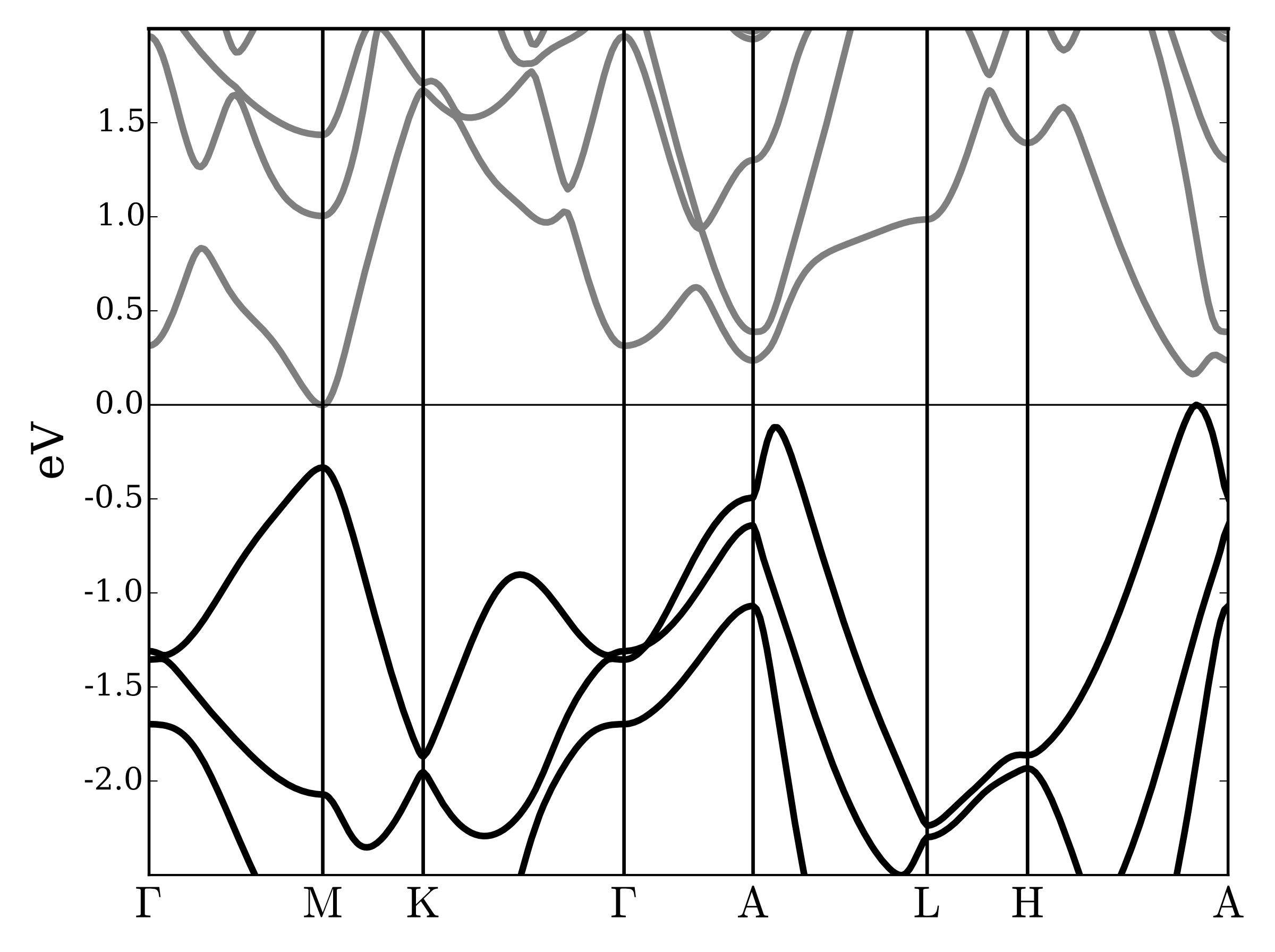}\label{fig:bulkgga}}
\subfigure[]{\includegraphics[bb= 0 0  576 430,width=8.5cm]{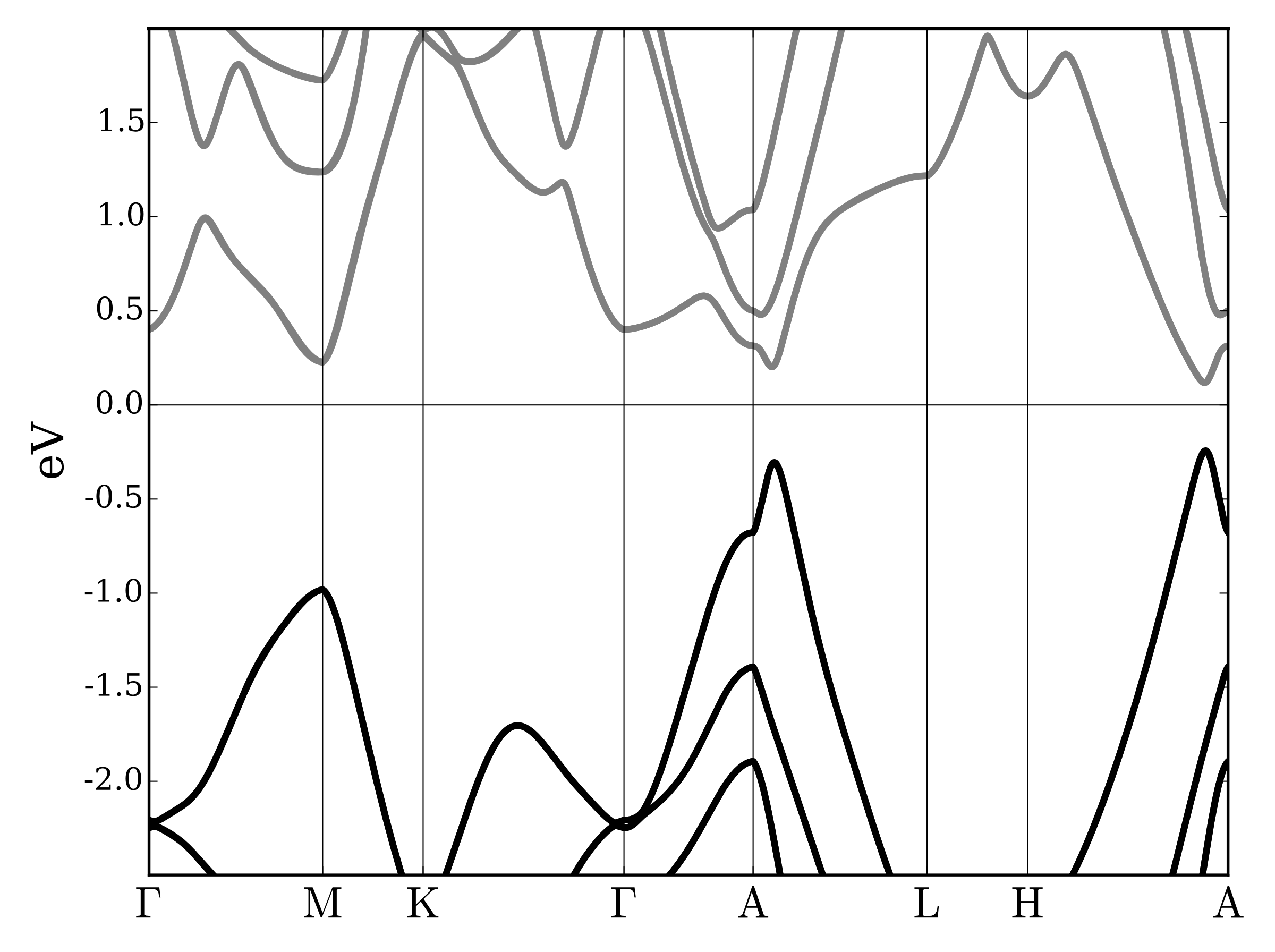}\label{fig:bulkhse}}

\caption{\subref{fig:bulkgga}~The bulk bandstructure of BaSn$_2$ computed using the PBE functional. The topological band inversion is seen to occur at A=(0,0,$\pi$), with spin-orbit splitting opening a gap of $\sim$200meV. However, the conduction band at M sits at the same energy as the HOMO maximum near A.  \subref{fig:bulkhse} The bulk bandstructure computed with the HSE hybrid functional.  The major effect of the inclusion of exact exchange is to widen the gap at M moving the conduction band clear of the Fermi energy.  Additionally, near A, the band gap is increased to $\sim$360meV. }\label{fig:bulkbands}
\end{figure}

BaSn$_2$ comprises alternating layers of tin atoms, arranged in a stanene-like buckled-honeycomb lattice~\cite{Xu_2013,Zhu_2015}, and barium atoms, set above/below the center of each plaquette of the tin layer.  The lattice parameters obtained at the level of PBE are $a=4.670$\AA, $c=5.450$\AA, in good agreement with experimental values and prior calculations~\cite{Kim_2008,ak31} (Table~\ref{tab:latt}).  The height of the tin layer is 1.15\AA. 
\begin{table}[h]
\begin{tabular}[c]{l | c | c | c}
\hline
                             & $a$(\AA) & $c$(\AA) & $z$(\AA)\\
\hline
Experiment~\cite{Kim_2008}   & 4.652 & 5.546 &  1.14\\
Experiment (this work)       & 4.655 & 5.515 & -\\
QE (PBE)                     & 4.670 & 5.450 & 1.15 \\
VASP (PBE)                   & 4.760 & 5.613 & 1.17\\
\hline
\end{tabular}
\caption{ The BaSn$_2$ lattice constants and tin layer height determined in previous experiments, the experiments described in this work, and the DFT calculations described in this work.}
\end{table}\label{tab:latt}

Fig.~\ref{fig:bulkbands} shows the band structures computed with both PBE and HSE functionals. It was shown previously~\cite{ak31} that the density of states near the band edge resides primarily in the covalent tin networks in each layer, so that the large barium atoms serve to control and mediate the weak but significant interlayer interactions evinced by the band structure.  With PBE, the band gap is indirect and almost exactly zero, with the conduction band at M falling to the Fermi energy. A topological band inversion, confirmed by calculation of the $Z_2$ invariant using parity eigenvalues~\cite{ak31}, is apparent at A=$(0,0,\pi)$, where pockets appear in the both the valence and conduction bands. Inversion at A rather than $\Gamma$ is unique; all other verified strong 3D topological insulators invert at $\Gamma$. Away from A the direct gap narrows to approximately 200meV.  
The HSE calculation reveals a significantly improved situation. The conduction band at M is shifted well-clear of the Fermi energy, so that the material is now an insulator.  Furthermore, the direct gap increases to approximately 360meV near A.   

\begin{figure}
{
\includegraphics[bb= 0 0  340 430, width=8.5cm]{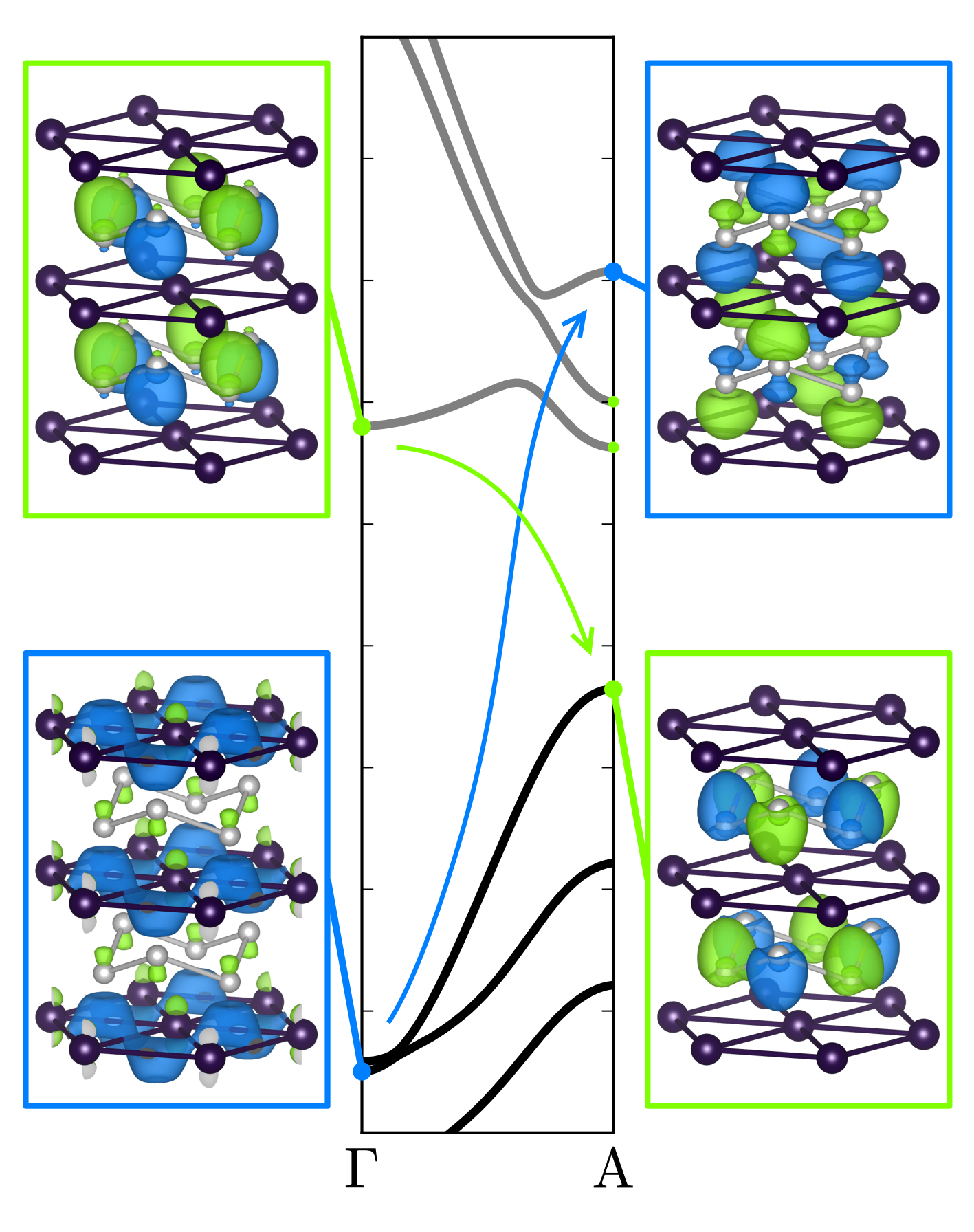}
}

\caption{In the center, the band structure along the line from $\Gamma$ to A is shown, with parity eigenvalues at the end points marked in blue ($+1$) and green ($-1$).  On either side, the wavefunctions are shown for the bands that invert.  The negative parity band, made of up tin $sp_z$ states lying mostly in the stanene layer, changes little along the line; however, the strong inversion of the positive parity band is seen to arise from the vanishing of the strong hybridization of tin and barium orbitals moving from $\Gamma$ to A.}\label{fig:inversion}
\end{figure}

At the point of inversion A, the band structure appears to exhibit the characteristic signature of inversion, where the effective masses of the HOMO and LUMO bands change sign.  However, more careful analysis reveals that these are not the relevant inverted bands.  The parity eigenvalues are both negative at A, indicating that this inversion is not the topological one.  The next even-parity band sits approximately 1eV above the Fermi level; it is this band that exchanges ordering with the HOMO at A.  This is reinforced by inspecting the wavefunctions directly at $\Gamma$ and A (Fig.~\ref{fig:inversion}); we see that the indicated valence(conduction) band at $\Gamma$ possesses the same orbital character as the marked conduction(valence) band at A. We note that the valence band at $\Gamma$ and the conduction band at A are both dominated by $sp_z$-like tin orbitals; however, while at $\Gamma$ these hybridize with Ba $sd_{z^2}$ orbitals, such hybridization is symmetry forbidden at A, where both the conduction and valence states indicate effectively isolated stanene layers. This suggests that we can understand the topological insulating state in BaSn$_2$ as originating from the interactions between stanene layers: the relevant bands in isolated stanene are inverted at $\Gamma$ of the 2D Brillouin zone, making it a quantum spin hall insulator.  A 3D system of widely separated, non-interacting layers will thus have the band inversions at $\Gamma$ and A. Then, if the barium and additional tin layers are brought into proximity, the interlayer interaction uninverts the bands at $\Gamma$, but is is forbidden to do so at A, resulting in a single band inversion in the 3D system and a strong topological index.

\subsection{Surface}
A direct consequence of the band inversion that occurs in the bulk is the appearance of metallic states localized to the interface between the topological insulator and a trivial insulator (such as vacuum).  They take the form of a Dirac cone pinned to a time-reversal invariant momentum that cannot be gapped without breaking time-reversal symmetry or closing the bulk gap. However, while the \emph{existence} of these states is guaranteed, their detailed characteristics depend on the surface structure and chemistry.  Unlike Bi$_2$Se$_3$~\cite{Hsieh_2009}, and similar to TlBiSe$_2$~\cite{Kuroda_2010,Singh_2016}, BaSn$_2$'s layers do not form weakly-interacting, quasi-independent units that suggest a natural cleavage plane. The barium and tin layers have nominal charges of $+2$ and $-2$, respectively; no simple cut will avoid uncompensated charge.  Surface termination will thus strongly impact the surface states, as the chemical environment of the surface layers are modified by the disruption of the bulk.  Furthermore,  real surfaces are complicated and environment dependent, even in the relatively simple case of bismuth selenide~\cite{Chen_2012,He_2013,Edmonds2014,Shokri_2015}.  However, this also allows for the possibility of deliberate manipulation. In conventional semiconductors, surfaces are often passivated by adsorption of additional species or formation of interfaces~\cite{winfried_1995}. In topological insulators, it has been suggested that surface states can also be controlled using adsorbates~\cite{Wang_2014} and heterogeneous surface layers~\cite{Chang_2015}.  Understanding the surface states of BaSn$_2$ will therefore require special consideration of the impact of chemical environment.

\begin{figure}
\subfigure[]{\includegraphics[bb=0 0 720 360,width=8.5cm]{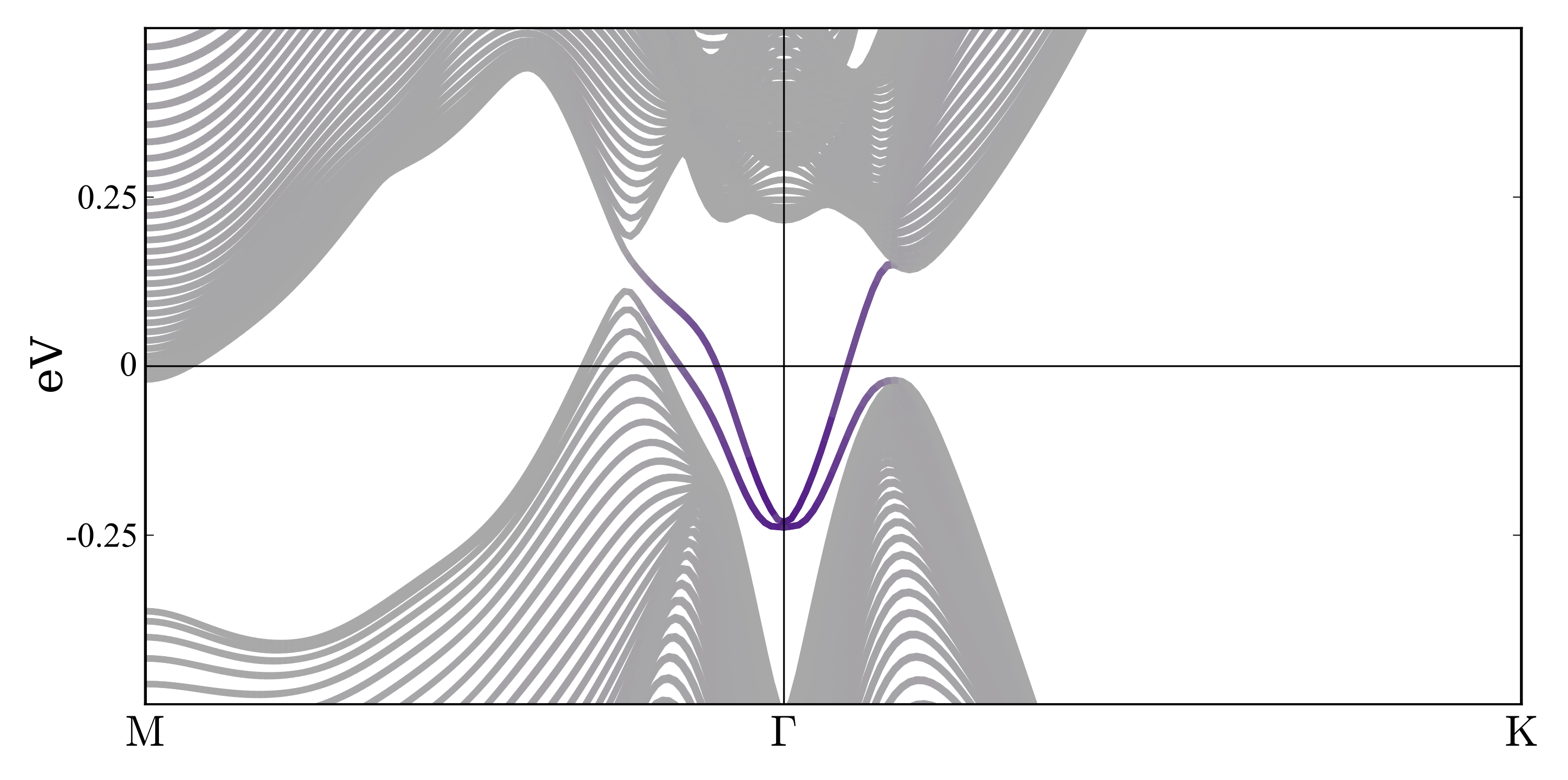}\label{fig:slabgga}}
\subfigure[]{\includegraphics[bb=0 0 720 360,width=8.5cm]{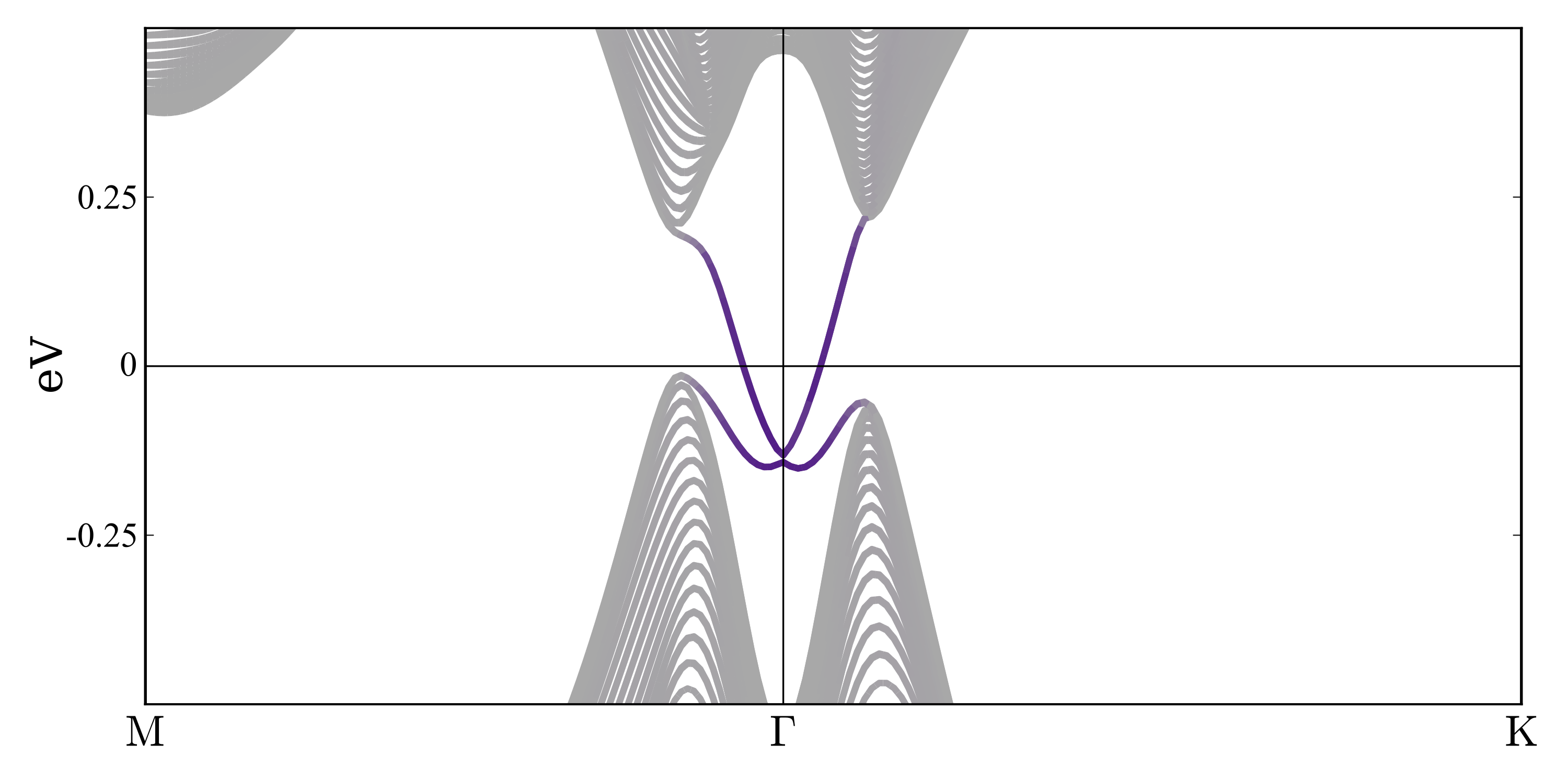}\label{fig:slabhse}}

\caption{Comparison of band structures for 50 layer slabs computed using the Wannier tight-binding model generated from~\subref{fig:slabgga} PBE and~\subref{fig:slabhse} HSE functionals.   The effect of HSE on the surface states is to increase the splitting between the conduction and valence bands, and lift the degeneracy at $\Gamma$ in energy, nearly into the bulk gap. Purple intensity denotes contribution of the Wannier functions localized to the top- and bottommost layers.}\label{fig:tb-slab}
\end{figure}

First, we compute the surface states using the tight-binding model, for which terminating the surfaces is straightforward, generated using the Wannier functions computed for the bulk.  We compare the PBE and HSE cases here as well.   
In Fig.~\ref{fig:tb-slab} the surfaces of 50 layer slabs are shown for both levels of theory. Ideally, well-defined Dirac cones would appear in the bulk gap at an energy uncrossed by other bands. In the present case, the topologically required surface states are clearly visible near $\Gamma$ and appear well-confined to the surface. However, they are afflicted by two issues.  First, there is weak splitting of the valence and conduction surface bands away from $\Gamma$, so that the Dirac cones are only just identifiable as such.  Second, the bands experience significant dispersion, driving them below the bulk gap, leaving the degeneracy at $\Gamma$ below the surface valence band maximum, as occurs at the surface of Bi$_2$Te$_3$~\cite{Chen2009}.  Both of these effects are substantially less severe in the model derived from the HSE calculations. 

\begin{figure*}
\subfigure[]{\includegraphics[bb=0 0 720 360,width=5.8cm]{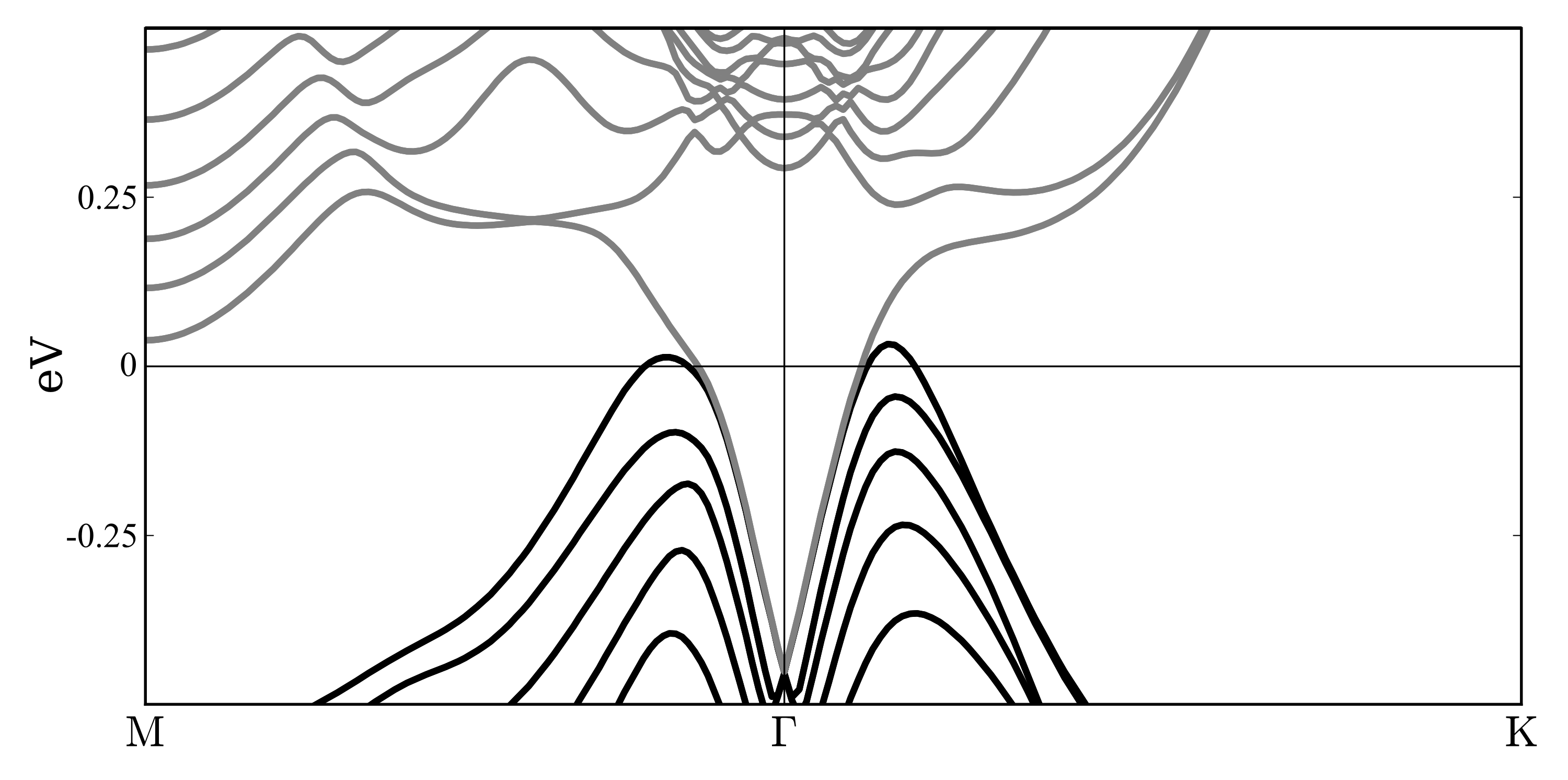}\label{fig:baslab}}
\subfigure[]{\includegraphics[bb=0 0 720 360,width=5.8cm]{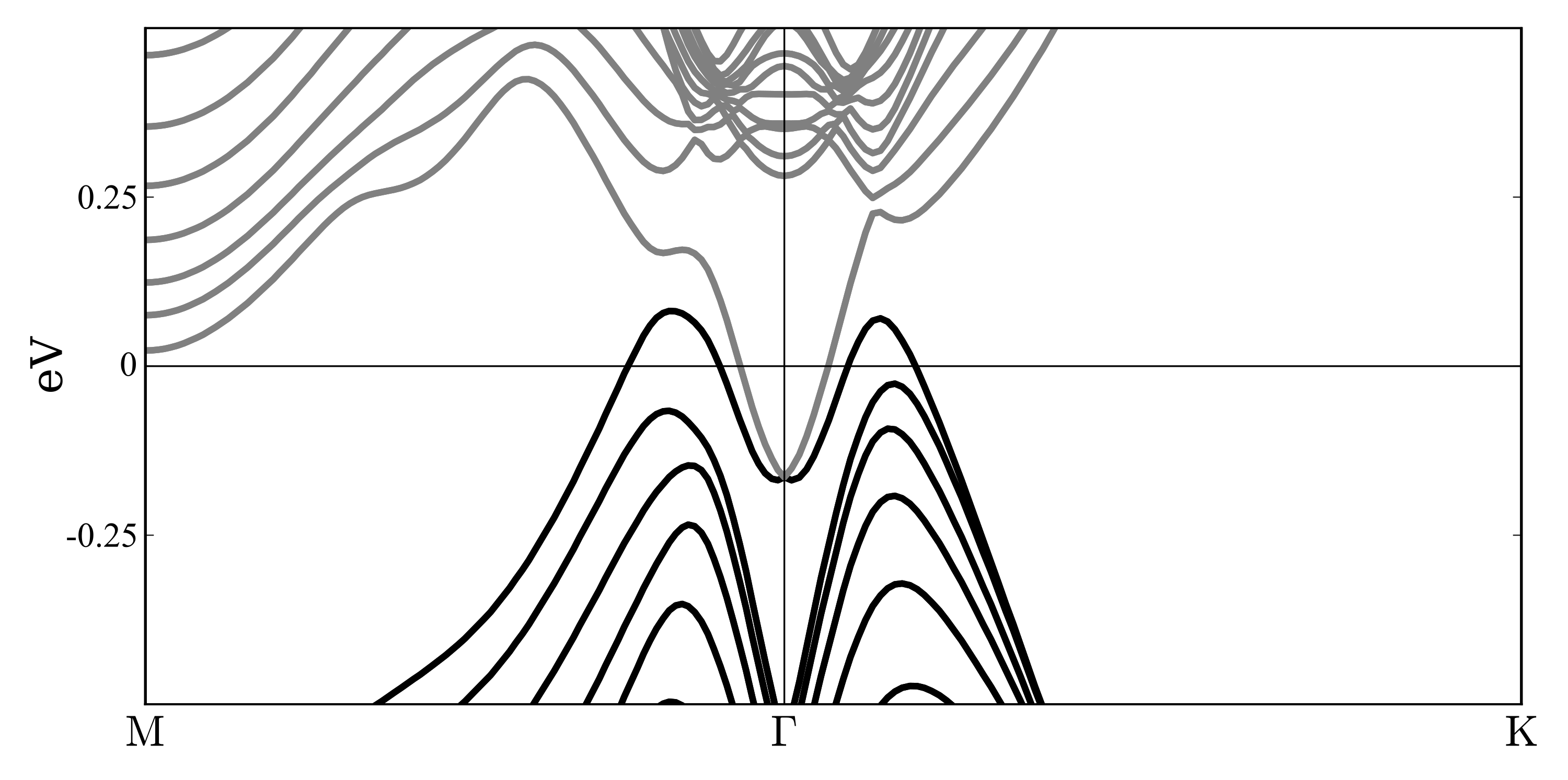}\label{fig:baclslab}}
\subfigure[]{\includegraphics[bb=0 0 720 360,width=5.8cm]{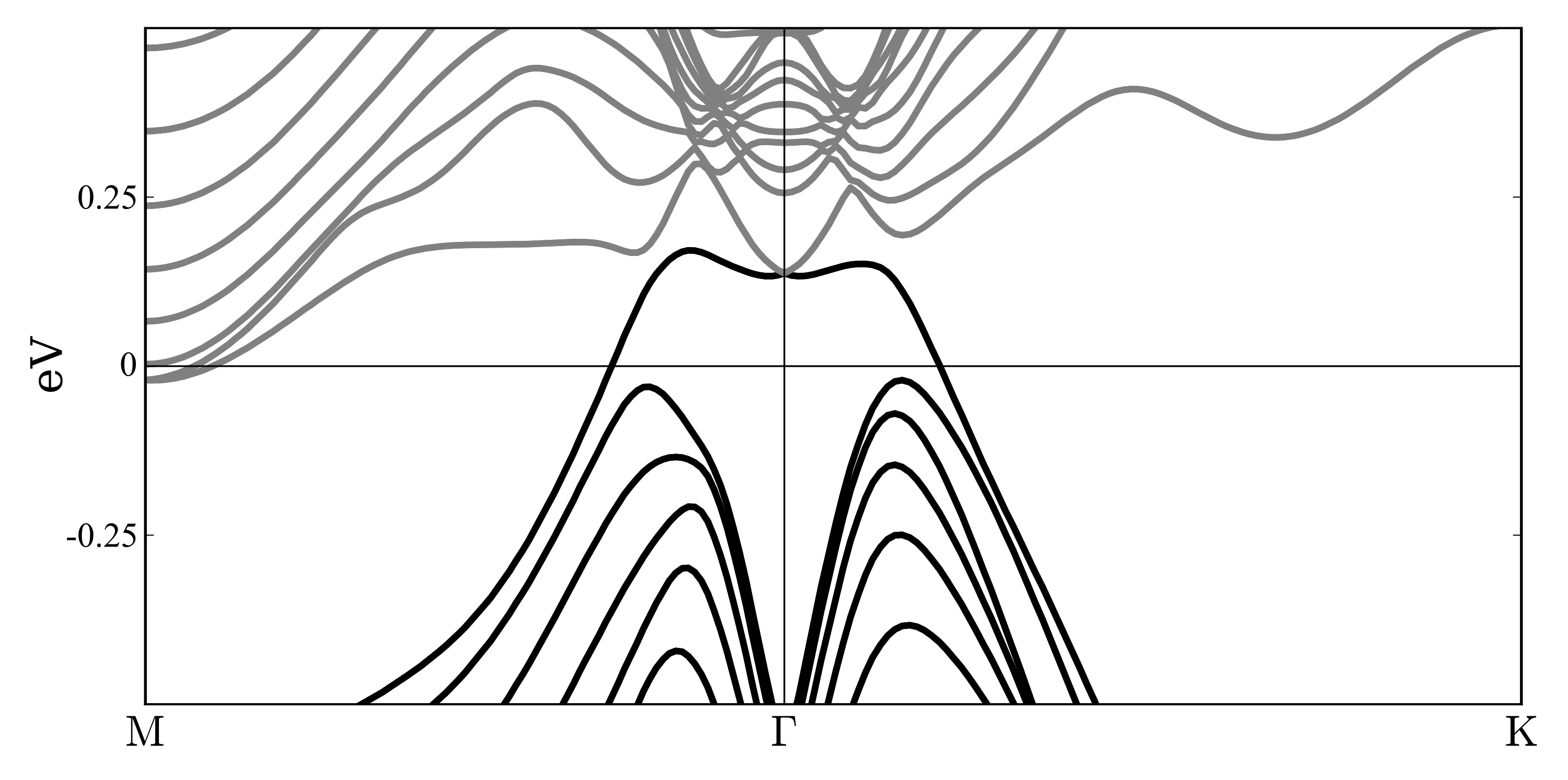}\label{fig:tlslab}}

\caption{PBE band structures for three different terminating compositions.~\subref{fig:baslab} Barium layer termination, but with potassium substituted for barium.  The resulting surface states lie deep within the valence manifold, and are nearly degenerate.~\subref{fig:baclslab} Inclusion of a monolayer of chlorine above a full layer of barium raises and splits the surface bands.~\subref{fig:tlslab} Substituting thalium for bismuth raises the surface bands even further, into the conduction manifold, while substantially increasing the surface band splitting. }\label{fig:dft-slab}
\end{figure*}

To investigate the impact of a real surface on the Dirac-cone surface states, we compute and examine the electronic structure of slabs with three simple terminations representing different chemical environments, shown in Fig.~\ref{fig:dft-slab}. 
In the first case, we consider termination at the barium layer. To avoid doping the system, either half-filling of the surface sites or substitution with an alkali metal, as has been proposed in MgB$_2$~\cite{Despoja2011}, is required. Calculations of both produce similar results; we present the latter, using potassium in place of barium, for ease of comparison. The resulting surface states are of very poor quality, magnifying the problems observed using the model surface. The surface states well within the valence bulk, and there is almost no splitting of the surface valence and conduction bands. 
In the second case, we consider a full-layer barium termination passivated by chlorine. The surface states are much improved, appearing very similar to the model calculation. We note that oxygen, as part of an oxide layer or hydroxide ion, may be expected to provided a similar chemical environment.  Finally, we terminate with a stanene layer, passivated by a layer of thallium.  This raises the degeneracy at $\Gamma$ into the conduction band, so that it is only slightly below the maximum of the surface valence band. The splitting of the surface bands is quite large as well. Supposing that the impact of the HSE functional observed in the model calculation represents physics that will also apply in the present cases, we consider it highly plausible that the true surface state can, in principle, be made to lie in the bulk gap at a Fermi energy uncrossed by states elsewhere in the Brillouin zone.

The above progression can be understood by considering the impact of outer-shell, non-bonding electrons in the vicinity of the topmost stanene layer.  Thallium has a shallow, filled $s$-orbital that significantly raises the energy of the proximal tin states.  Chlorine has a weaker impact, sitting above the barium layer, mimicking the effect of the tin that occupies the corresponding site in the bulk, and explaining the similarity between its resulting surface bands and those from the model.  Pure barium, however, has no shallow filled states, resulting in tin states lying very low in energy.  This suggests that the best results are likely to be obtained for stanene-terminated surfaces or interfaces including electron-rich species.
We stress that these calculations are primarily to demonstrate the variation possible in surface state quality depending on termination and surface treatment; identifying and optimizing realistic surface chemistries will require the investigation, both theoretically and experimentally, of significantly more configurations in terms of both composition and structure. 

\section{Stability and Synthesis}
BaSn$_2$ was originally obtained by fusing Ba and Sn at 1120 K
\cite{ak31,Ropp}. The long annealing at 770 K needed to isolate
BaSn$_2$ crystals and the presence of BaSn$_3$ byproduct indicate
unfavorable thermodynamics or kinetics governing the compound's
formation. Our following DFT analysis clarifies the thermodynamic
stability of bulk BaSn$_2$ under various $(P,T)$ conditions.

\begin{figure}[b]
\begin{center}
\includegraphics[width=87mm,angle=0]{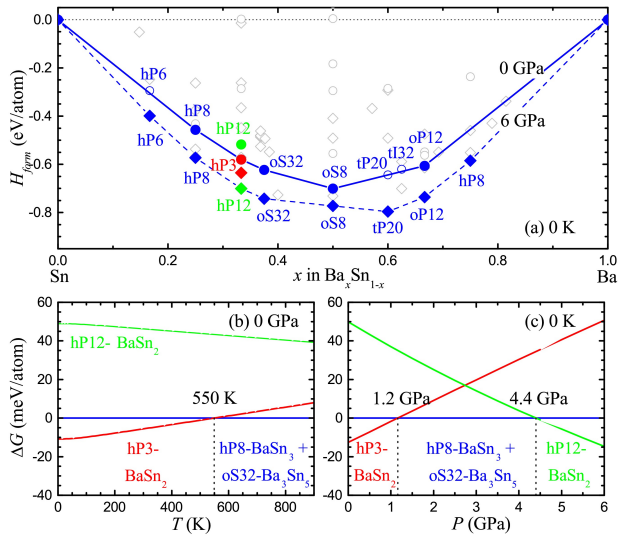}
\caption{ (Color online) (a) Calculated formation enthalpies for
  relevant Ba-Sn phases at $T=0$ K. Circles and diamonds show the
  results at 0 GPa and 6 GPa, respectively. As described in the text,
  the shown hP12-BaSn$_2$ phase could be a preferred starting material
  for obtaining crystalline hP3-BaSn$_2$. (b,c) Calculated Gibbs
  energies of hP3-BaSn$_2$ and hP12-BaSn$_2$ as a function of $T$ or
  $P$ relative to the combination of hP8-BaSn$_3$ and
  oS32-Ba$_3$Sn$_5$.}\label{fig:phases1}
\end{center}
\end{figure}

Formation energy results at $T=0$ K and $P=0$ GPa in Fig.~\ref{fig:phases1}(a) are
consistent with the reported observations of (meta)stable Ba-Sn
compounds at the 1:5, 1:3, 1:2, 3:5, 1:1, 3:2, 5:3, and 2:1
stoichiometries~\cite{Ropp}. Considering that hP3-BaSn$_2$ is found
only 12 meV/atom below the tie-line connecting neighboring
hP8-BaSn$_3$ and oS32-Ba$_3$Sn$_5$, the compound's stability could be
easily affected by the synthesis conditions. Based on the
Gibbs energy dependence in Fig.~\ref{fig:phases1}(b), hP3-BaSn$_2$ is indeed
destabilized at temperatures above 550 K. In order to have the
possibility to grow single crystals out of liquid the transition
temperature should be at least close to the liquidus line which is,
unfortunately, difficult to establish in this case. Experimentally,
the liquidus line around this Sn-rich stoichiometry has yet to be
mapped out~\cite{Ropp}. Computationally, the transition temperature
estimates in DFT approximations are prone to large systematic errors
\cite{ak31,ak30}; e.g., our local density approximation
\cite{LDA1,LDA2} calculations indicate that hP3-BaSn$_2$ is already
unstable by 14 meV/atom at $T=0$ K.

The conclusions of this stability analysis were supported by our crystal growth attempts.  We used initial compositions with Ba:Sn ratios of 37:63, 33:67, and 30:70.  In all cases the final product, after being cooled and spun down, was polycrystalline Ba$_3$Sn$_5$(space group: $Cmcm$, no.: 63)~\cite{Ropp} chunks and BaSn$_3$ (space group: $P63/mmc$, no.: 194)~\cite{massalski_1990,Ropp} rod-like crystals. This demonstrates that the published Ba-Sn phase diagram is incomplete~\cite{massalski_1990}, as no Ba$_3$Sn$_5$ phase is indicated; our results show a clear signature of this phase towards the Ba-rich side exposed to the Ba-Sn liquidus line.  
No signature of BaSn$_2$ was observed, which suggests that near the stoichiometry range of BaSn$_2$, this phase is not connected to the Ba-Sn liquidous line. Hence, it does not seem possible to grow BaSn$_2$ single crystals from Ba-Sn liquid by flux method. 
\begin{figure}
\begin{center}
\includegraphics[width=8.5cm]{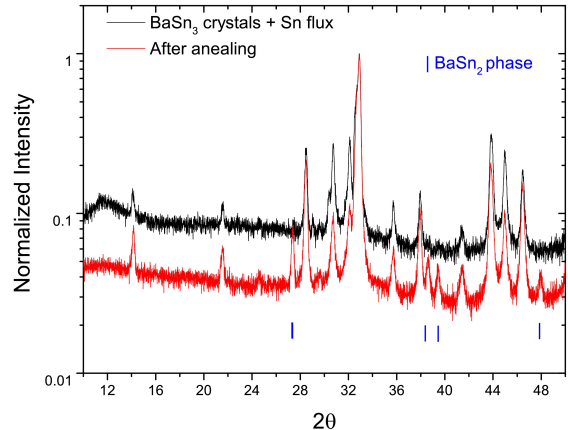}
\caption{Powder XRD of BaSn$_3$ crystals and Sn flux obtained from Ba$_{30}$Sn$_{70}$ growth (black line) and powder XRD after low temperature annealing showing appearence of BaSn$_2$ phase (red line). }\label{fig:pxrd}
\end{center}
\end{figure}

We next annealed BaSn$_3$ crystals and Sn flux at 500$^\circ$~C for 7 days, after which clear signatures of BaSn$_2$ were observed.  Fig.~\ref{fig:pxrd} shows powder X-ray diffraction (PXRD) for BaSn$_3$ crysals and Sn flux before and after the annealing.  The new peaks that have appeared post-annealing at $2\theta=27.3^\circ,38.7^\circ,39.4^\circ$ and $48.0^\circ$, corresponding to (011), (110), (102), and (201) planes of the BaSn$_2$ phase, respectively.  This indicates that BaSn$_2$ is a low temperature phase hidden beneath BaSn$_3$ phase and forms via slow low temperature decomposition of the BaSn$_3$ phase.

\begin{figure*}[t]
\begin{center}
\includegraphics[width=180mm,angle=0]{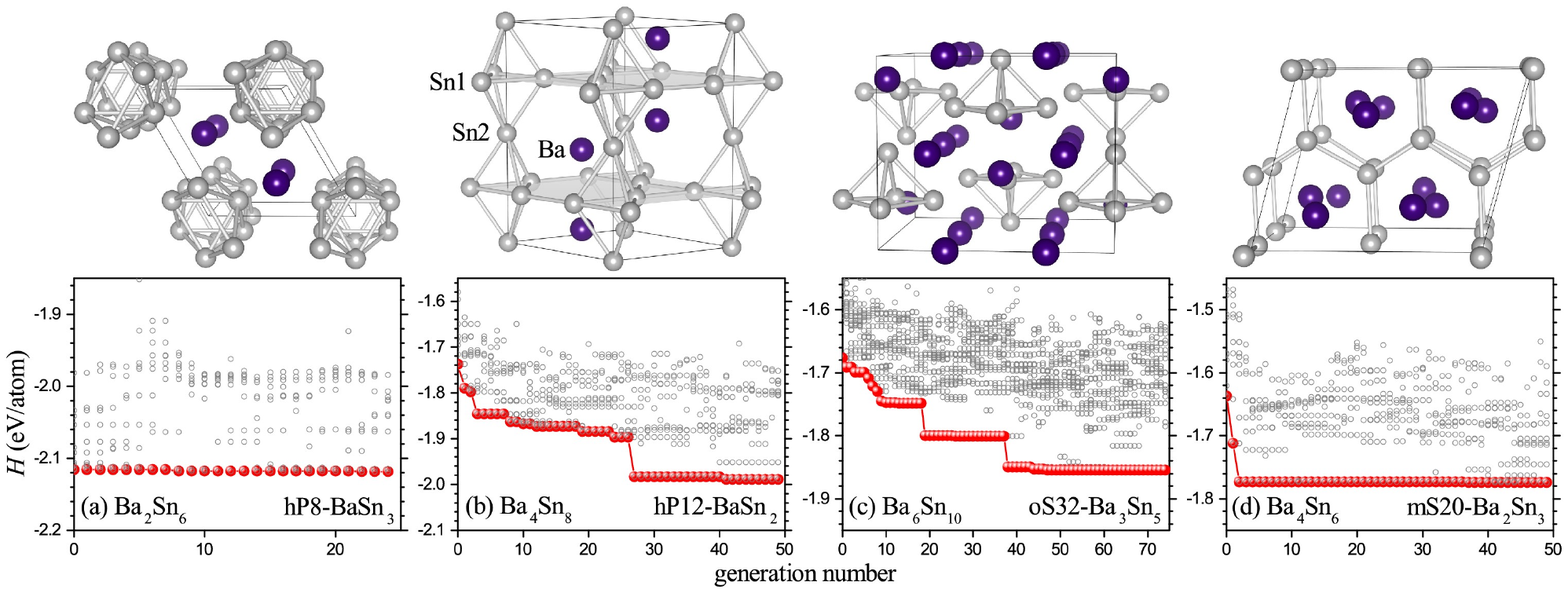}
\caption{ (Color online) Distributions of enthalpies during
  evolutionary structure optimization runs at $P=10$ GPa for unit
  cells with 2:6, 4:8, 6:10, and 4:6 Ba and Sn atoms. The best member
  in each generation is shown in red. The structures shown above each
  panel were found to have the lowest enthalpy at the corresponding
  composition. In the hP12-BaSn$_2$ phase, expected to be stable
  above $\sim$4 GPa, the Sn framework intercalated by Ba atoms
  comprises the kagome lattice bridged by Sn2 atoms.}\label{fig:phases2}
\end{center}
\end{figure*}

A common approach to stabilizing certain phases only marginally stable
under ambient conditions is application of hydrostatic pressure
\cite{ak16,ak31}. With a large measured interlayer distance of 5.515
\AA, hP3-BaSn$_2$ happens to have a relatively large volume per atom
(36.4 \AA$^3$/atom) compared to the average one (32.9 \AA$^3$/atom)
for the hP8-BaSn$_3$ and oS32-Ba$_3$Sn$_5$ mixture. As a result, the
enthalpic $PV$ term quickly destabilizes hP3-BaSn$_2$ (Fig.~\ref{fig:phases1}(c)). Interestingly, a new BaSn$_2$ phase with the Laves hP12
structure observed in the K-Sn system becomes stable at an estimated
4.4 GPa pressure readily achievable in multi-anvil cells
(Fig.~\ref{fig:phases2}). Our phonon calculations indicate that hP12 is
dynamically stable at both 0 and 10 GPa. The vibrational entropy
contribution lowers the relative Gibbs energy of hP12 (Fig.~\ref{fig:phases1}(b)) and
should reduce the transition pressure by $\sim 0.5$ GPa at synthesis
temperatures around 600 K.

Based on this finding, an unconventional two-step route can be
suggested for obtaining the desired hP3 polymorph. If hP12 does form
at high pressures and can be quenched down to ambient conditions it
might be a more suitable starting material for producing crystalline
hP3. Thermodynamically, the hP12 phase is found to be at least 30
meV/atom above hP3 in the 0-800 K range at $0$ GPa (Fig.~\ref{fig:phases1}(b)).
Kinetically, the hP12$\rightarrow$hP3 composition-preserving
transformation is likely to have lower barriers, compared to those in
the hP8-BaSn$_3$+oS32-Ba$_3$Sn$_5$$\rightarrow$hP3-BaSn$_2$
solid-solid reaction. The desired rearrangement of the 3D Sn framework
in hP12 (Fig.~\ref{fig:phases2}(b)) would involve migration of the bridging Sn2 atoms
into the kagome lattice comprised of the Sn1 atoms to form the
denser-packed but wider-separated hexagonal Sn layers.

In order to check whether any other phases could prevent the formation
of hP3-BaSn$_2$ or hP12-BaSn$_2$ in the discussed pressure range, we
performed evolutionary ground state structure searches at the 1:3, 1:2, 3:5 and 2:3
compositions at 10 GPa. For each unit cell size specified in Fig.~\ref{fig:phases2},
we did 2-3 separate runs starting with randomly generated structures.
The populations of 10-16 members were evolved over 25-75 generations
with crossover operations (70\%) or mutation operations (30\%)
\cite{ak23,ak31}. Typical enthalpy profiles in Fig.~\ref{fig:phases2} illustrate good
diversity maintained in the evolved populations as similar structures
are eliminated from the pools. The results indicate that the
ambient-pressure hP8-BaSn$_3$ and oS32-Ba$_3$Sn$_5$ remain
lowest-enthalpy structures at these compositions up to 10 GPa. At the
1:2 stoichiometry, the searches converged to the hP12 structure
discussed above. Best 2:3 structures with up to 2 formula units were
found to be metastable by at least 20 meV/atom at 0 GPa and 10 GPa.
According to these calculations, it seems possible to obtain
hP3-BaSn$_2$ via the proposed high-pressure route.

\section{Conclusion}
We have presented a new strong topological insulator, BaSn$_2$, composed of alternating stanene and barium layers.  HSE calculations reveal a bulk gap of 300-400meV, one of largest gaps of any known topological insulator.  The surface states are strongly dependent on termination and surface chemistry. Our results demonstrate that high quality surface states are possible though not given.  Attempts at synthesis confirm that BaSn$_2$ is a stable, low-temperature phase, but rule out direct growth of single crystals from liquid flux.  Additional calculations suggest an alternate route to single crystals may be possible via a two-step, high-pressure synthesis.  However, despite these challenges, we feel the unique composition and large bulk gap make BaSn$_2$ a worthwhile target for further investigation.

\section{Acknowledgments}
While this paper was being reviewed, we became aware of Ref.~\cite{Huang_2016}, which considers BaX$_2$ with X=Sn,Ge,Si as potential topological nodal semimetals (i.e., in the absence of spin-orbit coupling).

S. M. Y. was supported by a
National Research Council Research Associateship Award at the US Naval
Research Laboratory. J.S. and A.N.K. acknowledge the NSF support (Award No. DMR-1410514). The work in Ames was supported by the U.S. Department of Energy, Office of Basic Energy Science, Division of Materials Sciences and Engineering. The research was performed at the Ames Laboratory. Ames Laboratory is operated for the U.S.
Department of Energy by Iowa State University under Contract No. DE-AC02-07CH11358. In addition, S.M. was supported by the Gordon and Betty Moore Foundations EPiQS Initiative through Grant GBMF4411.

\end{document}